\documentclass[a4paper,11pt]{article}
\pdfoutput=1 

\usepackage{jinstpub} 

\usepackage{subcaption}
\usepackage{color,soul}
\usepackage{lineno}

\title{\boldmath Development of a real-time tritium-in-water monitor}

\author[a,1]{C.D.R.~Azevedo,\note{Corresponding author.}}
\author[b,2]{A.~Baeza\note{Deceased}}
\author[c]{E.~Chauveau}
\author[d]{J.A.~Corbacho}
\author[e]{J.~D\'iaz}
\author[c]{J.~Domange}
\author[c]{C.~Marquet}
\author[e]{M.~Mart\'inez-Roig}
\author[c]{F.~Piquemal}
\author[a]{D.~Prado}
\author[a]{J.F.C.A.~Veloso}
\author[e]{N.~Yahlali}

\affiliation[a]{I3N, Physics Department, University of Aveiro,\\Campus Universit\'ario de Santiago, 3810-193 Aveiro, Portugal}
\affiliation[b]{Universidad de Extremadura, Laboratorio de Radioactividad Ambiental,\\ Servicio de apoyo a la investigaci\'on, C\'aceres, Spain}
\affiliation[c]{University of Bordeaux, CNRS/IN2P3/CENBG,\\ UMR 5797, F-33170 Gradignan, France}
\affiliation[d]{Departamento de Física Aplicada, Centro Universitario de Mérida, Universidad de Extremadura,\\ 06800 - Mérida (Badajoz), Spain.}
\affiliation[e]{Departamento de Física Aplicada, Centro Universitario de Mérida, Universidad de Extremadura,\\ 06800 - Mérida (Badajoz), Spain.}
\emailAdd{cdazevedo@ua.pt}

\abstract{In this paper, we report the development and performance of a detector module envisaging a tritium-in-water real-time activity monitor. The monitor is based on modular detection units whose number can be chosen according to the required sensitivity. The full system is being designed to achieve a \emph{Minimum Detectable Activity} (\emph{MDA}) of 100\,Bq/L of tritium-in-water activity which is the limit established by the E.U. Council Directive 2013/51/Euratom for water intended for human consumption. The same system can be used as a real-time pre-alert system for nuclear power plant regarding tritium-in water environmental surveillance. The first detector module was characterized, commissioned and installed immediately after the discharge channel of the Arrocampo dam (Almaraz nuclear power plant, Spain) on the Tagus river. Due to the high sensitivity of the single detection modules, the system requires radioactive background mitigation techniques through the use of active and passive shielding.  We have extrapolated a \emph{MDA} of 3.6\,kBq/L for a single module being this value limited by the cosmic background. The obtained value for a single module is already compatible with a real-time environmental surveillance and pre-alert system. Further optimization of the single-module sensitivity will imply the reduction of the number of modules and the cost of the detector system.}


\dedicated{This paper is dedicated to the memory of Professor Antonio Baeza (1955 - 2022), the greatest mentor and leader of the TRITIUM project.}

\arxivnumber{2202.11068} 

\begin{document}
\maketitle
\flushbottom

\section{Introduction}
\label{sec:intro}

The E.U. Council Directive 2013/51/Euratom \cite{directive2013} establishes a limit of 100\,Bq/L of tritium in water intended for human consumption. For the measurement of tritium concentrations within the limit defined by the legislation, the usual method is the Liquid Scintillation Counting technique \cite{Parker2023}, which can take up to 3\,days from the sample collection till the result. Moreover, a real-time tritium low activity concentration measurement can be used as a nuclear power plant environmental surveillance and anomaly pre-alert system, for example, in the case of primary coolant leak \cite{Rathnakaran2000}. Presently and to the best of our knowledge, there is no such instrument with the required capabilities aforementioned. A concise review of tritium detection and assessment in aqueous media can be found in \cite{Parker2023}\\
The challenge of the tritium detection in water medium is due to the low energy $\beta$ emitted particles (5.7\,keV on average with a maximum of 18.6\,keV). Such low energy $\beta$ emissions in water presents a very low range (<5\,$\mu$m) \cite{Rathnakaran2000, Berthold1999} which implies the detection system sensitivity depends on the sensing area in contact with the water. In a previous simulation work, we have shown the possibility to use uncladded plastic scintillating fibers for the sensing of tritium in water. We chose fibers which maximize the surface in contact with water. Furthermore, removing the fibers clad allows to increase the betas interaction rate with the sensing core. A modular approach was conceived in order to provide scalability to the detection system, according to the required sensitivity \cite{Azevedo2020}. For low tritium concentration measurement and correct tritium decay identification, the presence of the natural and anthropogenic backgrounds must be efficiently suppressed. For the air-born and soil background a lead shield must be used, while for the radioisotopes dissolved in water a cleaning system is required. This water cleaning system must remove all algae and sediments from water (avoiding their deposition on the surface of the fibers) and de-ionize the water to remove all ionic and radioactive contaminants. Tritium will not be removed from the water through filtration as its dominant form is the tritiated water (HTO and T2O) \cite{Hou2018}. For this project, a water cleaning system that produces de-ionized water with conductivity of the order of 10\,$\mu$S/cm was built \cite{Azevedo2020_CP}. The water cleaning systems design, physicochemical parameters and radionuclide activities before and after the cleaning process can be found in \cite{corbacho}. Another source of natural background is the cosmic-rays that, despite the higher energy deposition in the fibers, can create lower energetic particles in the vicinity of the detector, mainly due to interaction with the lead shield, resulting in energy deposition in the region-of-interest \cite{Azevedo2020_CP}. To remove the cosmic particle signals a cosmic veto must be employed in anti-coincidence with the tritium detection modules.\\
In this work, we report on the experimental development and measurements of a single module prototype based on PMTs' light-readout. A report on a similar setup based on SiPMs is being prepared and will be published soon.

\subsection{The prototype setup}
\begin{figure}[tb]
	\centering
		\includegraphics[width=0.75\columnwidth]{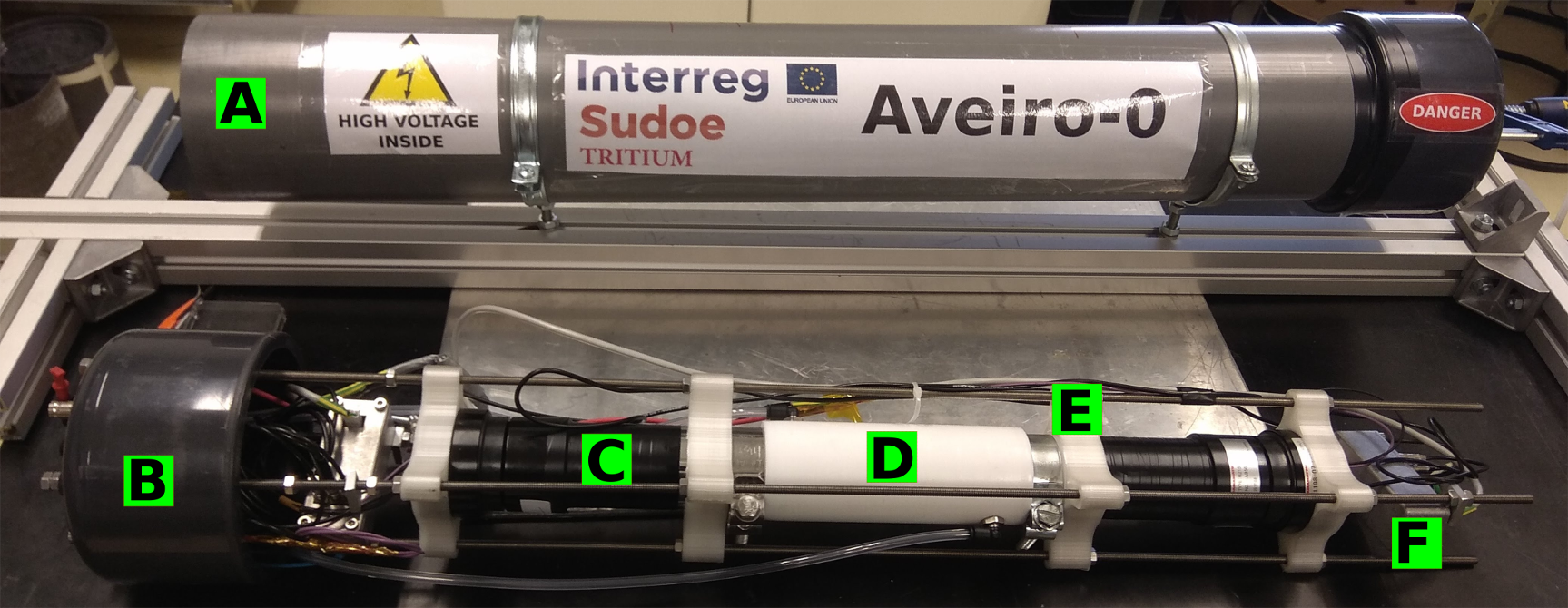}
	\caption{Picture of the TRITIUM Aveiro-0 prototype outside its PVC enclosure.}
	\label{FIG:1}
\end{figure}

\begin{figure}[htb]
\centering
\begin{subfigure}{.49\textwidth}
  \centering
  \includegraphics[width=1.\linewidth]{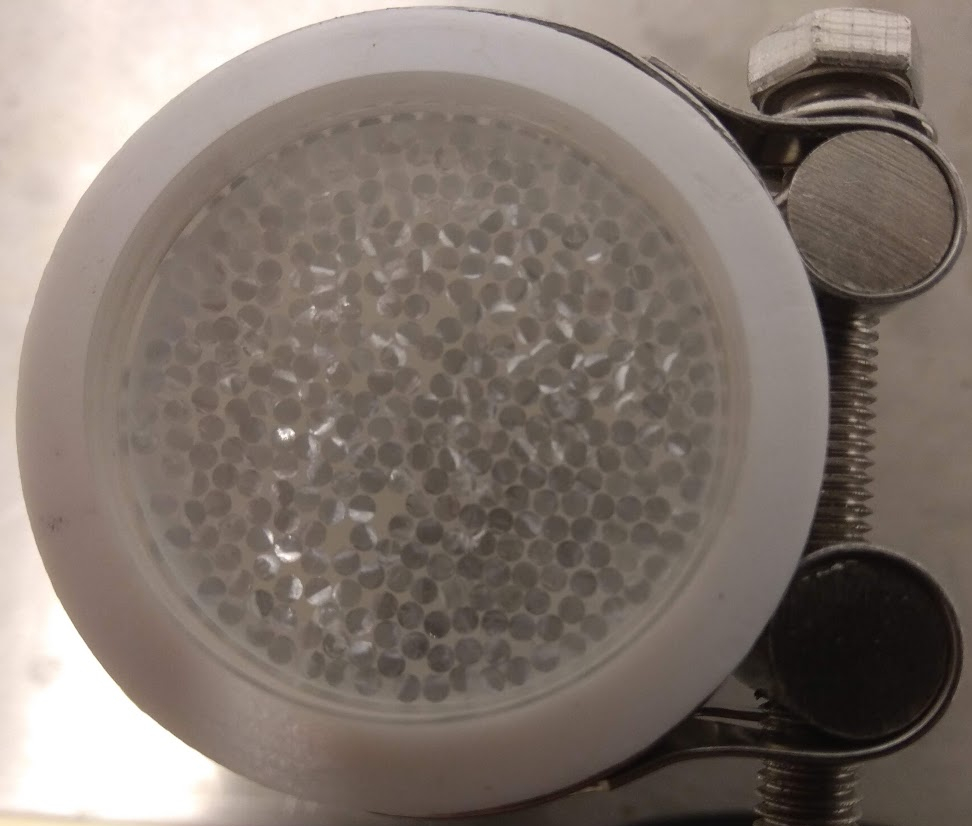}  
  \caption{Front side illumination}
  \label{FIG:2a}
\end{subfigure}\hfill
\begin{subfigure}{.49\textwidth}
  \centering
  \includegraphics[width=\linewidth]{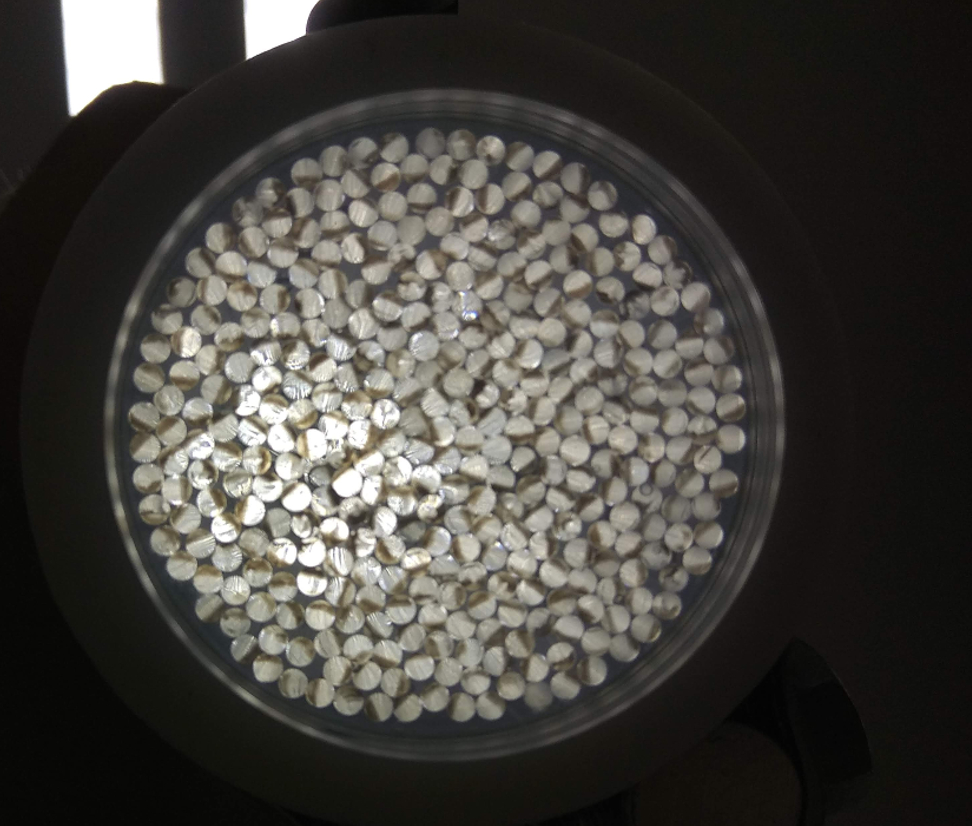}  
  \caption{Back side illumination}
  \label{FIG:2b}
\end{subfigure}
\caption{Pictures of the scintillating fibers inside the teflon cell}
\label{FIG:2}
\end{figure}

Figure \ref{FIG:1} is a picture of the so-called "Aveiro-0" prototype module partially disassembled. In the front plane, the inner instrumentation is shown while in the back plane the protective outer shell can be observed. The prototype was designed for easy access and fast assembly/disassembly in case of future maintenance. The outer shell is made from a PVC tube (\emph{A}) which protects the PMTs (\emph{C}) and the teflon (PTFE) tube (\emph{D}) from physical damage while providing a light-tight operation environment. The prototype is mounted by gently sliding the inner part into the PVC tube closing it with the PVC end-cap (\emph{B}). The PVC cap holds feed-through connectors for the PMTs' high voltage, pre-amplifiers low voltage and signals . The inner part and its components are assembled and held in place by 4 stainless-steel long screws attached to the PVC endcap. The positioning and centering of the PMTs in relation to the teflon tube is performed by 3D-printed supports (\emph{E}). To decrease the signal loss and noise, the pre-amplifiers are housed in 2 aluminium boxes (\emph{F}) positioned as close as possible to the PMTs.
The 20\,cm long teflon tube (\emph{D}) is closed by acrylic (PMMA) optical windows. The PMMA windows are fixed by radially pressing the teflon tube using clamps, which also provide water sealing, as shown in Figure \ref{FIG:2a}. The PTFE tube contains 360 uncladded scintillating fibers (18\,cm-long and 2\,mm-thick Saint-Gobain BCF-10 \cite{SaintGobain}) and an inlet/outlet for the water circulation. In Figure \ref{FIG:2b}, a picture of the module with back-light illumination shows the fiber cut and cleavage patterns. The fibers were not polished after the cut as they will act as a scintillator rather than light guides. For the light-readout two 2" Hamamatsu R2154-02 PMTs \cite{R2154} were used.

\subsection{Electronics chain}

\begin{figure}[tb]
\centering
\includegraphics[width=.9\textwidth]{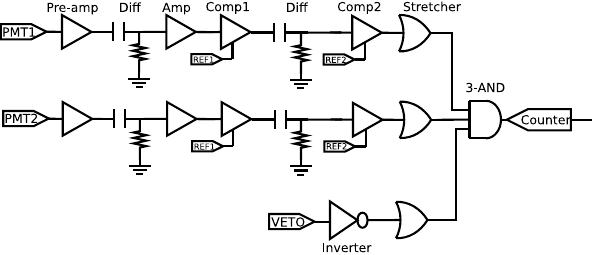}
\caption{Electronic simplified scheme for each single module readout and coincidence architecture.}
\label{FIG:3}
\end{figure}

Figure \ref{FIG:3} presents a simplified scheme of the electronic chain developed within this work. It is composed of 3 lines, two of which being identical and operated in coincidence in order to reject the PMTs photocathodes self-emission signals. The third line is the input for the anti-coincidence veto signal from the cosmic particles discriminator system.
  Each PMT signal is shaped and pre-amplified by a CREMAT CR111 pre-amplifier \cite{cremat} followed by differentiation and amplification stages. The amplifier is based on an OPA656 opamp \cite{opa656}. The amplified signal is fed to a fast comparator (LT111 \cite{LT111}) to apply a low-level threshold and reject low amplitude noise signals. The threshold levels of the comparators are established by a MAX5500 DAC \cite{MAX5500}. Due to the long decay time of the pre-amplifiers signals, the comparators output signals have durations of the order of 200\,$\mu$s which increases the false coincidence probability between the PMTs. To decrease the duration of the pre-amplifier signal, a second differentiation stage is used. As the comparator produces a 5\,V square pulse the differentiation of the edge will result in a faster signal with fixed amplitude. Next, a second comparator stage is used to produce again a 5\,V square pulse. A tunable pulse stretcher based on an OR gate (SN74AHC1 \cite{orGate}) is used to to set the duration of each PMT signal (100\,ns) resulting in a maximum time acceptance window of 200\,ns.
  The veto signal, assumed to be also a 5\,V positive square pulse, is inverted in order to perform anti-coincidence with the PMTs. Due to the inversion stage, the veto signal will be always high level except when a cosmic particle is detected becoming a low level. The signals of the PMTs plus the veto are logically level-compared by a 3-input AND gate (SN74LVC1G11 \citep{andGate}) with the output connected to a pulse counter. For the system control and threshold level evaluation, the output of each comparator is counted by a Raspberry Pi using the GPIO pins and interrupt routines. The output of the pre-amplifier is divided and fed to a voltage follower circuit whose output can be used for energy/signal amplitude studies.
  
\begin{figure}[tb]
  \centering
	\begin{subfigure}{.49\textwidth}
	\centering
  	\includegraphics[width=\linewidth]{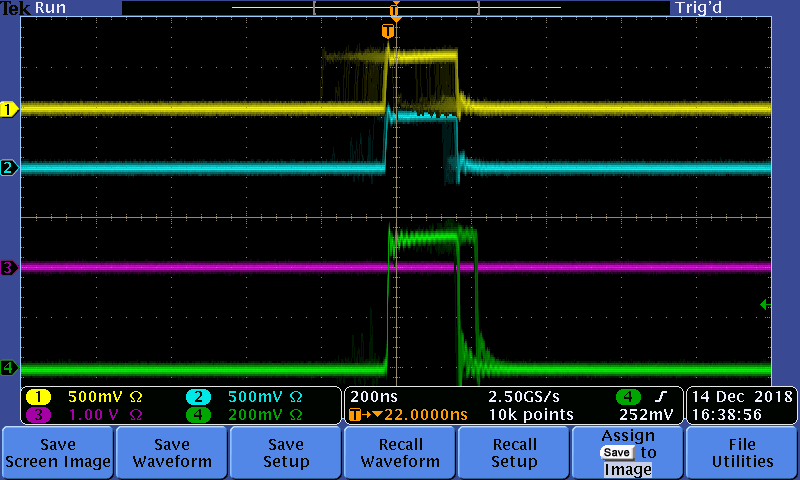}  
  	\caption{Coincidence true. No cosmic particle detected}
  \label{FIG:4a}
\end{subfigure}\hfill
\begin{subfigure}{.49\textwidth}
  \centering
  \includegraphics[width=\linewidth]{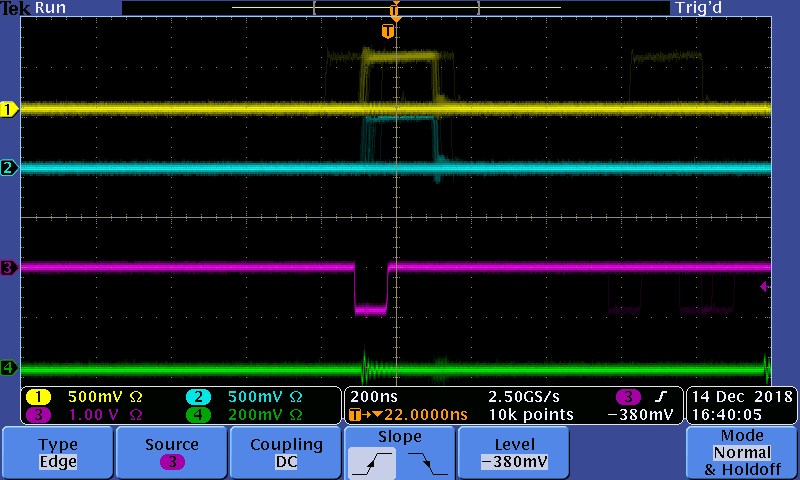}  
  \caption{Coincidence false. Cosmic particle detected}
  \label{FIG:4b}
\end{subfigure}
\caption{Example of typical signals from the coincidence circuit obtained during the single module commissioning.}
\label{FIG:4}
\end{figure}

In order to illustrate the PMTs' signals coincidence and veto anti-coincidence, two oscillogramms are shown in Figure \ref{FIG:4}. A single 10$\times$10\,cm\textsuperscript{2} scintillator was placed outside the PVC tube in the region above the fibers to produce a cosmic veto signal. The yellow (1) and cyan signals (2) are the output of the pulse stretchers of each single PMT. The pink signal (3) is the inverted veto while the green (4) is the 3-AND gate output. In Figure \ref{FIG:4a} two in-time signals from the PMTs without a signal from the veto are observed resulting in the triggering of the AND gate. In this case a true signal is counted. In Figure \ref{FIG:4b}, two coincident signals from the PMTs and a veto signal, corresponding to the interaction of a cosmic ray both in the veto and the fibers, are observed. In this case, no signal from the AND coincidence is triggered and so, the event is rejected.

The PMTs are biased at their maximum voltage (-1500V) in order to maximize the signal-to-noise ratio. The negative high voltages were sourced by two Hamamatsu C11152-01 \cite{C11152} controlled by a second MAX5500 DAC. For the DAC communication and cross-check of the output values, an Arduino Mega is used. All the communication, as well as the prototype slow control, were performed by the Raspberry Pi. 

\section{Characterization and measurements}
\subsection{Characterization of PMTs}

\begin{figure}[tb]
\centering
\begin{subfigure}{.49\textwidth}
  \centering
  \includegraphics[width=\linewidth]{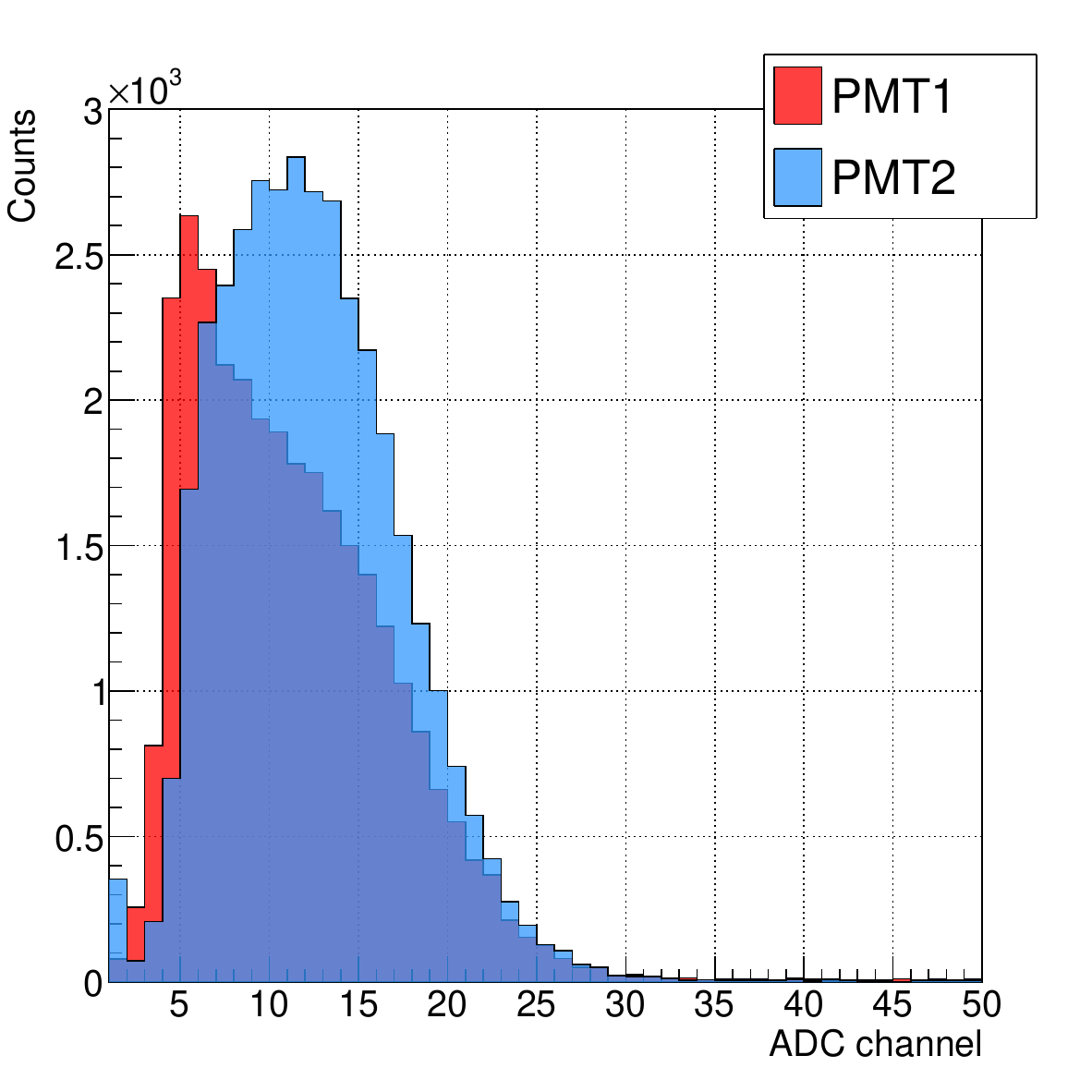}  
  \caption{Self trigger, PMTs windows closed}
  \label{FIG:6a}
\end{subfigure}\hfill
\begin{subfigure}{.49\textwidth}
  \centering
  \includegraphics[width=1.\linewidth]{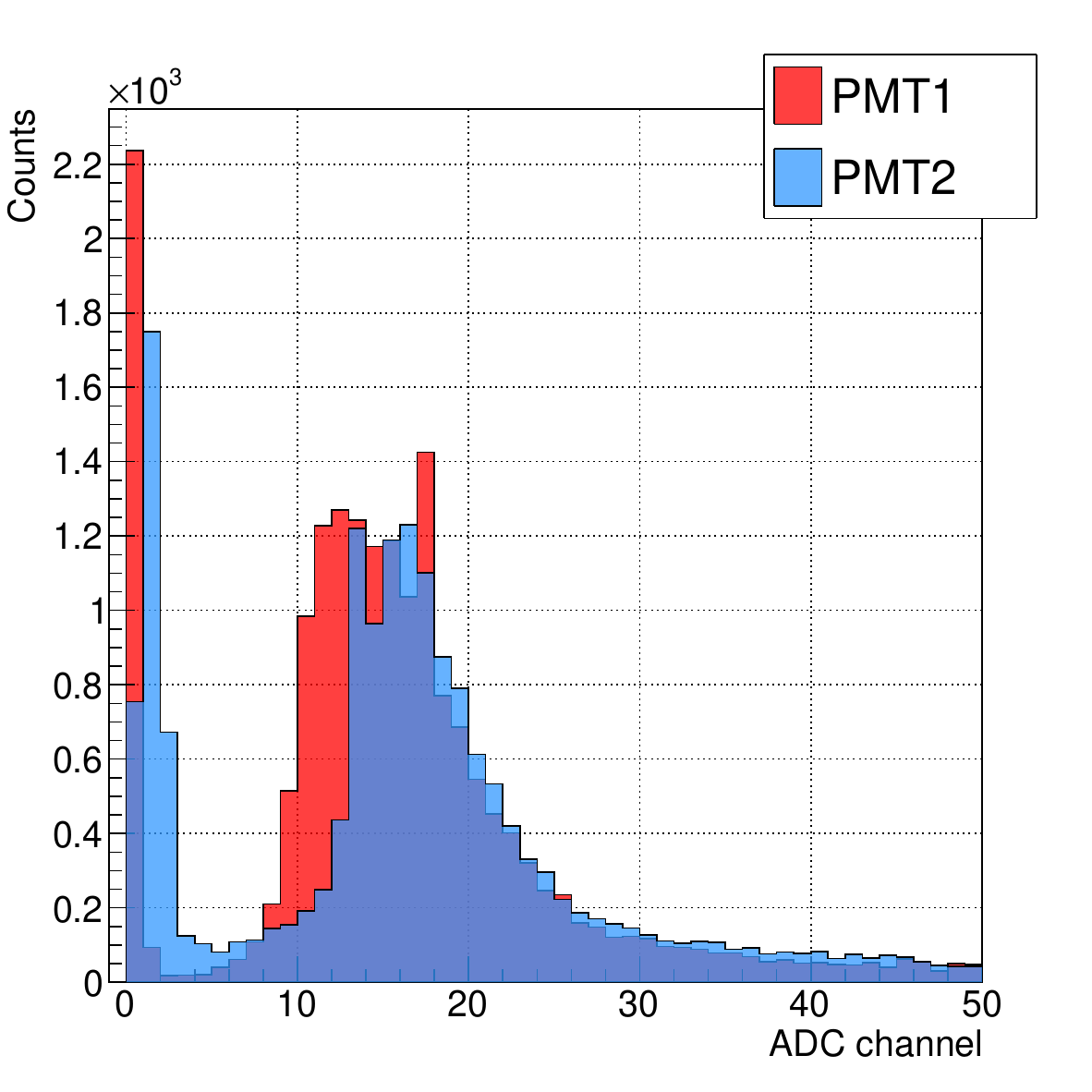}  
  \caption{External trigger, PMTs windows open}
  \label{FIG:6b}
\end{subfigure}
\caption{Pulse-height energy distribution for each single PMT.}
\label{FIG:6}
\end{figure}

In order to characterize the PMTs single-photon energy distribution, the signals from the charge-sensitive pre-amplifiers have been digitized, shaped and the pulse height measured by a CAEN V1724 digitizer in self-trigger mode. To obtain a single-photon energy distribution the teflon cell was removed and the PMTs windows were covered with black caps in order to use the self emitted photoelectrons from the photocathode. Figure \ref{FIG:6a} presents the single-photoelectron energy distribution of both PMTs where the effect of the applied thresholds in the digitizer is observed through the cut on the lower energy channels. The PMT1 single-electron peak is not deconvoluted from the noise due to a lower signal-to-noise ratio of this PMT relatively to PMT2. Aiming to characterize the energy of events that can produce fake coincidences, a second acquisition with the PMTs windows uncovered and in coincidence was performed.  The signals from the PMTs were fed to the coincidence board to produce an external trigger for the digitizer in case of timing coincidence.  The results are presented in Figure \ref{FIG:6b} where the peaks positions are compatible with the single-photoelectron energy distribution (Figure \ref{FIG:6a}). The number of observed false coincidences (without fibers) indicates a light leak through the feedthroughs. An observable difference between both figures is the higher population after the peaks in the case when the windows are open. Such higher energy events can be explained by the Cherenkov production in the PMTs glass by cosmic particles and the natural glass radioactivity \cite{knoll_PMTS,electronTubes}. The results of this measurement may indicate a probable need of low radioactive background PMTs to achieve the sensitivity goal of 100\,Bq/L.\\

\subsubsection{Prototype tests in coincidence mode}
\begin{figure}[tb]
\centering
\begin{subfigure}{.49\textwidth}
   \centering
   \includegraphics[width=\linewidth]{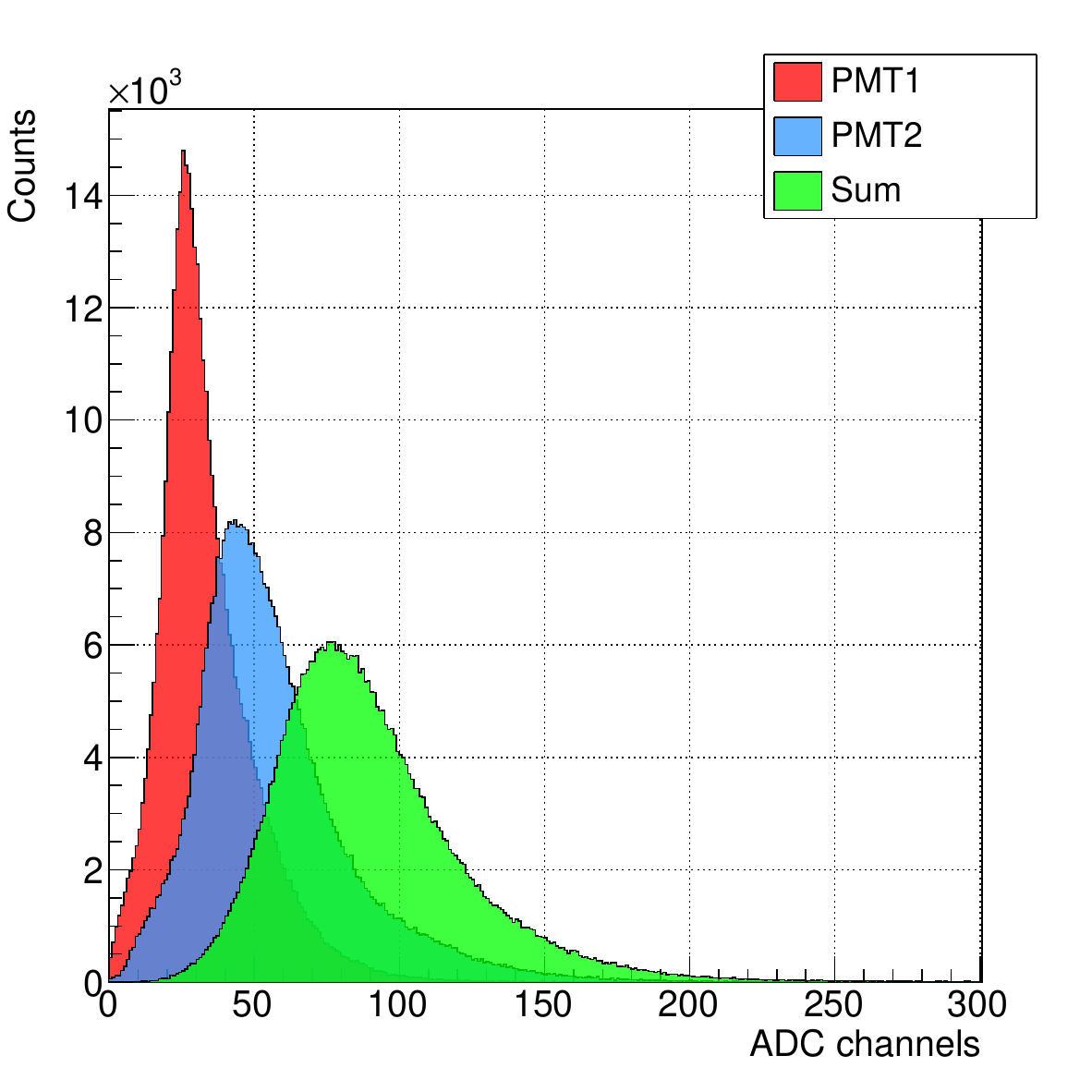}  
   \caption{Experimental results }
   \label{FIG:7a}
 \end{subfigure}\hfill
 \begin{subfigure}{.49\textwidth}
   \centering
   \includegraphics[width=1.\linewidth]{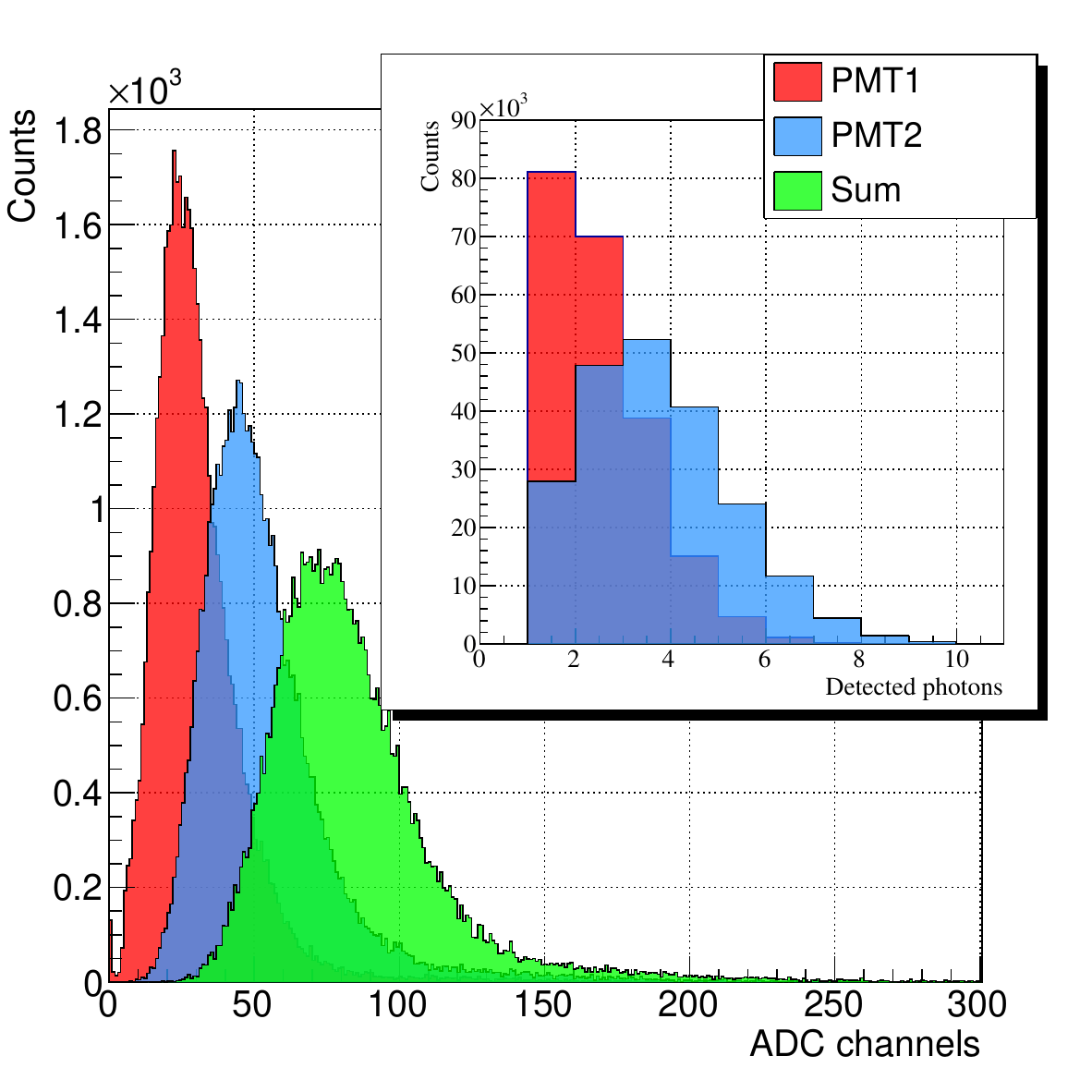}  
      \caption{Simulation}
   \label{FIG:7b}
 \end{subfigure}
\caption{Pulse-height distribution for each single PMT and their sum using a  \textsuperscript{55}Fe source ($\gamma$-5.9\,keV) located close to the teflon tube extremity. a) Experimental results. b) Simulation. The inset in this figure is the simulated number of detected photons per event that produces a coincidence.}
\label{FIG:7}
\end{figure}

The teflon tube containing the fibers was positioned and coupled to the PMTs using Saint-Gobain BC-630 silicone grease. In order to characterize the prototype using coincidence mode, a $\gamma$ source was used due to its energy peak for the sake of results interpretation. A \textsuperscript{55}Fe source was chosen$E_{\gamma} = $5.9\,keV) which is close to the energy deposition peak for a in-water tritium source ($\sim$5\,keV according to the work presented in \cite{Azevedo2020}). Due to the strong attenuation by the materials for the $5.9$\,keV $\gamma$ photons, the source had to be placed inside the teflon tube in between the fibers and close to one of the windows. Owing to the presence of the source inside the sensing module, the measurement was performed without water. The acquisition scheme was similar to the previous measurement but this time using an external trigger produced by the described electronics circuit. With the external trigger, just signals in time coincidence will be digitized (coincidence mode). This technique allows to remove the influence of the electronic noise and the signals produced by the PMTs self-emission. In Figure \ref{FIG:7a} the acquired pulse-height distributions of each PMT and their sum is presented. It is observed that PMT2 presents its peak at higher values relative to PMT1. This effect is explained by a higher PMT gain and by the source position which was located close to it, resulting in a lower photon attenuation relatively to this PMT. The second observation that must be addressed is the tail observed in each PMT distribution that is also present in the energy sum histogram. In order to understand the pulse-height distribution a simulation of the 5.9\,keV $\gamma$ photons was performed using the GEANT4 code described in \cite{Azevedo2020}. The previous code was changed so that the fibers were surrounded by air, the source positioned between the fibers and close to the respective PMT. The PMTs signals were simulated by a uniform sampling of the experimental single-electron distributions presented in Figure \ref{FIG:6a}. For each simulated $\gamma$ photon detection, the number of produced optical photons was computed. The signal of each optical photon was obtained by a random sample taken from the measured distributions and summed for all the optical photons in a single event. In Figure \ref{FIG:7b} the results of the study are presented where we can observe a good agreement between the shape of the experimental and simulated distributions. The simulated distributions present lower standard deviations that can be related to the absence of electronic noise and residual light as shown in Figure \ref{FIG:6}, that was not included in the simulation. The effect of the source position is observed on the relative peaks' position of each PMT. In the simulated results the tail on the right side of the peaks is also observed. This tail is explained by the distribution of detected photons per event presented in the inset of Figure \ref{FIG:7b}. 

\subsection{Counting mode}
Due to the characteristic continuous $\beta$  emission energy  distribution, the continuous energy loss of $\beta$ particles (resulting in a detection energy dependence with the emitted position), the signal amplitude dependence with the interaction position, and the fluctuation in the number of produced photons (see ref. \cite{Azevedo2020}), the pulse-height distribution will result in a broad continuous were useful information is difficult to extract. In this scenario, just the number of coincidences will be used by counting them, the so-called counting mode. From this point, all the reported data was acquired using the counting mode.

\subsection{Passive shield}

\begin{figure}[tb]
	\centering
		\includegraphics[width=0.5\columnwidth]{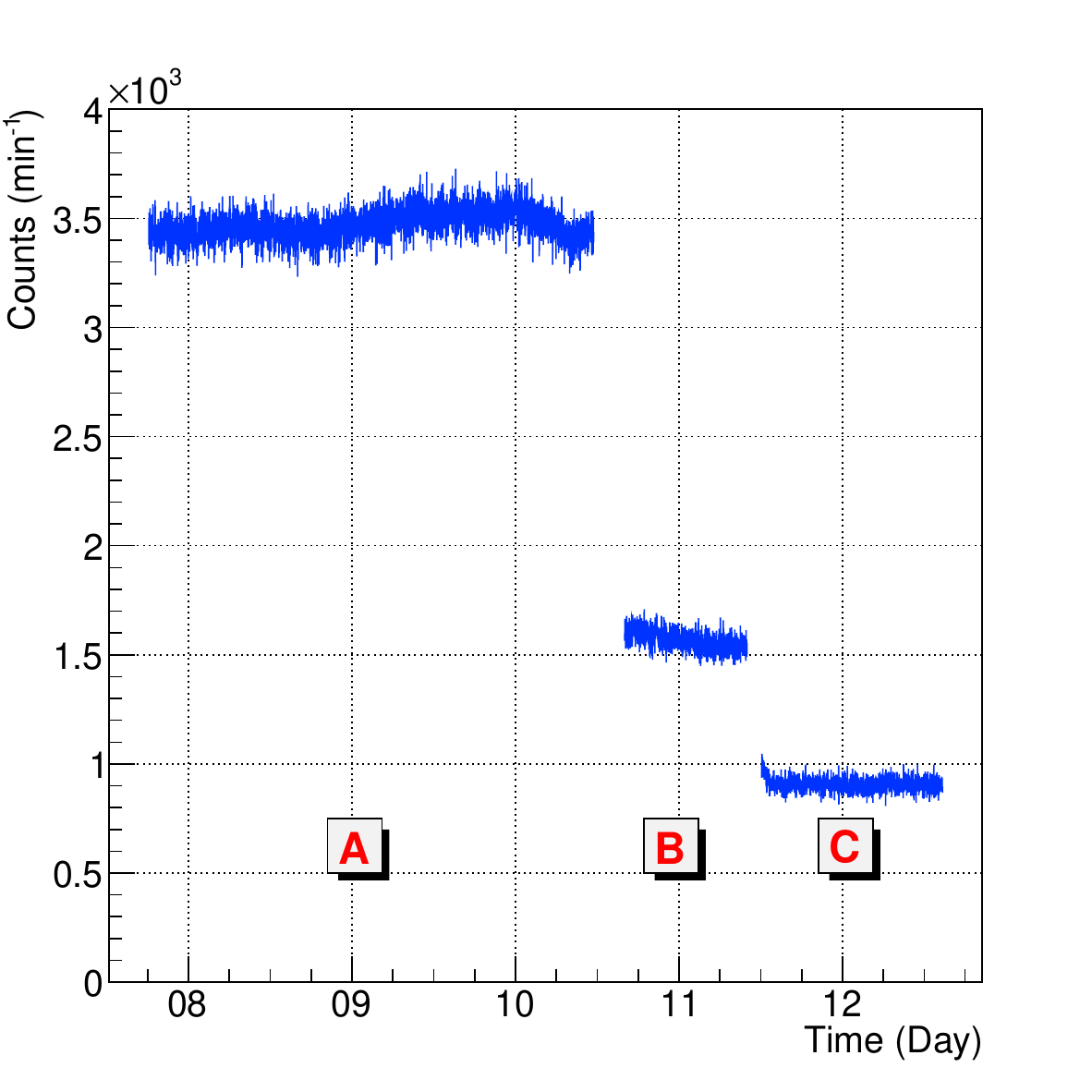}
	\caption{Counting rate for different lead thicknesses wrapped around the sensing module: A - No lead, B - 2.5\,mm lead, C - 5\,mm lead. The measurements were performed without water in the module.}
	\label{FIG:9}
\end{figure}

In order to evaluate the radioactive background, the \textsuperscript{55}Fe source was removed and the acquisition system changed for counting mode. In Figure \ref{FIG:9}, region $A$, the data acquired during 2.5\,days are shown, with an average value of $3.5 \times 10^{3}$\,counts/min. To visualize the effect of the passive shield in the background reduction a lead foil (2.5\,mm thickness) was wrapped around the PVC tube (region $B$). A background reduction of more than a factor of 2 was measured. Another lead sheet was added resulting in a lead thickness of 5\,mm, which allows reducing the background level of about a factor 4 (region $C$) relatively to the initial condition where no passive shield was used.

\subsection{Prototype commissioning: Minimum Detectable Activity}

\begin{figure}[tb]
\centering
\begin{subfigure}{.49\textwidth}
  \centering
  \includegraphics[width=\linewidth]{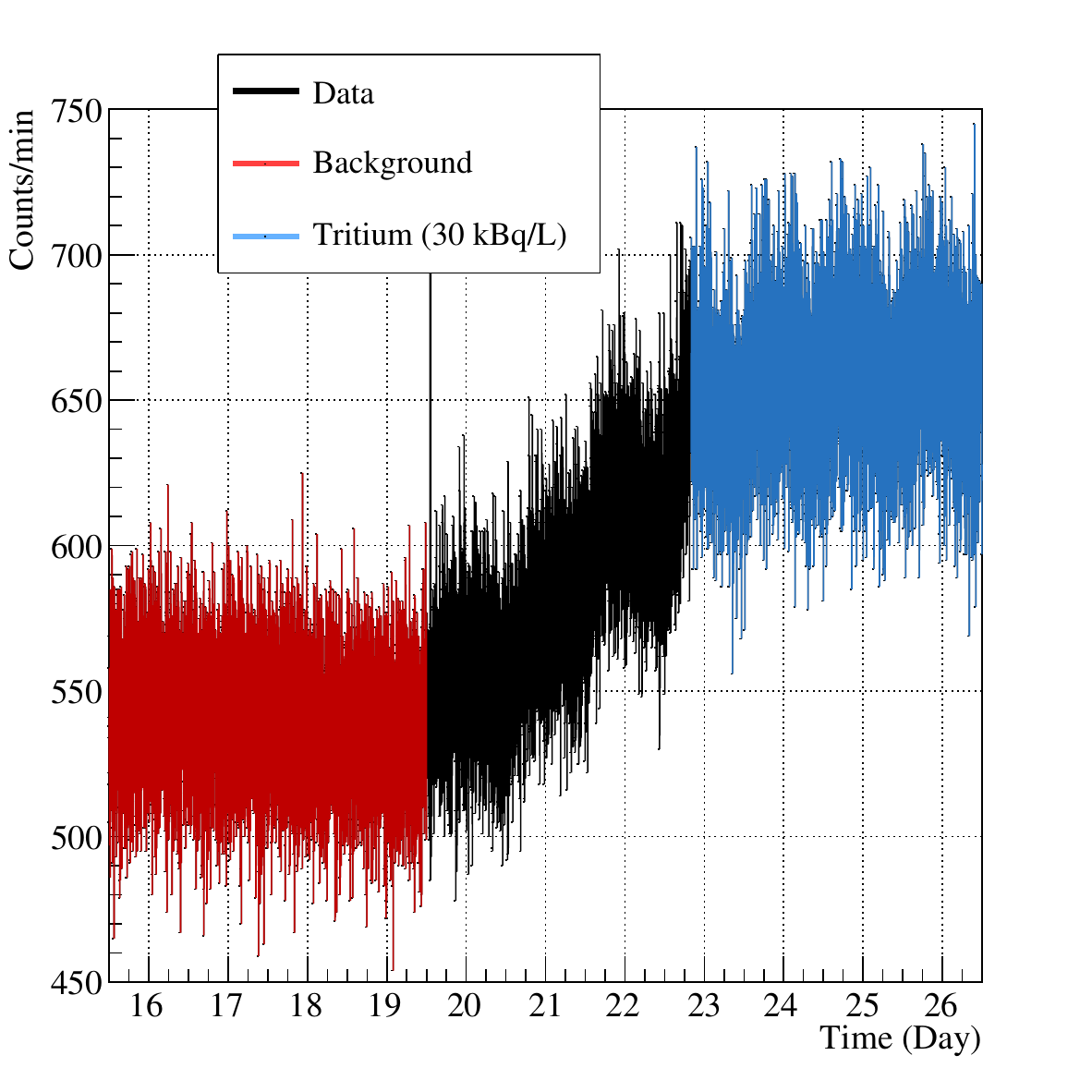}  
  \caption{Counting rate acquired in real-time measurement }
  \label{FIG:10a}
\end{subfigure}\hfill
\begin{subfigure}{.49\textwidth}
  \centering
  \includegraphics[width=\linewidth]{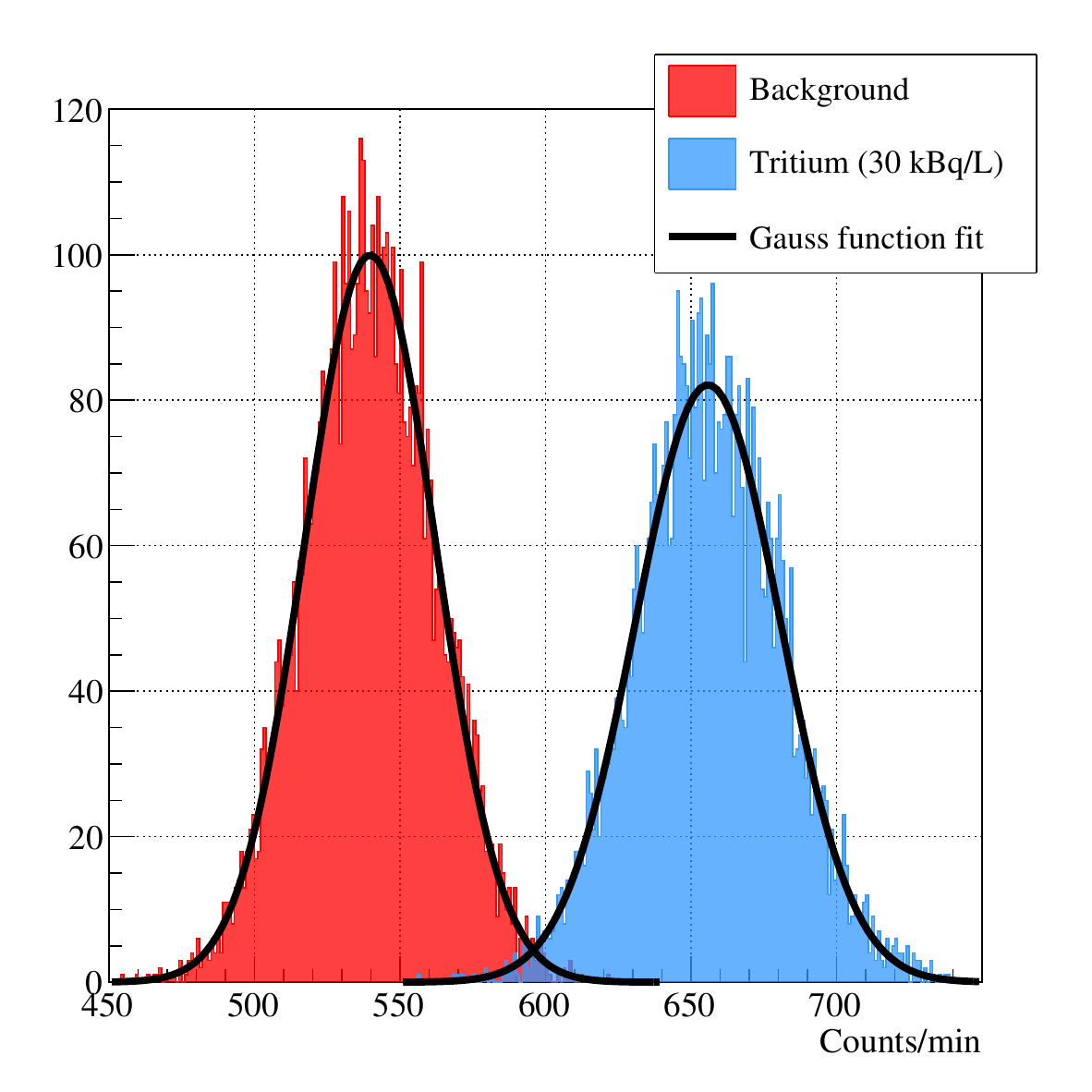}  
  \caption{Distribution of the selected data for background and real tritium activity (30\,kBq/L)}
  \label{FIG:10b}
\end{subfigure}
\caption{Background and tritium concentration measured in a 1\,min integration time by increasing the tritium concentrations till 30\,kBq/L.}
\label{FIG:10}
\end{figure}

Prior to the prototype installation, the system was moved and commissioned at the laboratories of the University of Extremadura (Spain). The prototype was positioned horizontally and surrounded by 5\,cm thick lead bricks. The prototype was filled with pure water and a background measurement was performed during 4\,days.The data are represented by the red line in Figure \ref{FIG:10a}. It was obtained an average background level ($N_B$) of 540\,counts/min with a standard deviation ($\sigma_{Nb}$) of 23\,counts/min. The distribution of the background counting rate is represented by the red histogram in Figure \ref{FIG:10b}. The quality of the Gaussian fit allows to state that the statistical fluctuations are mainly due to the counting statistics. The Minimum Detectable Activity ($MDA$), i.e., the minimum net counts that ensures a false-negative rate no larger than 5\% ($N_D$) when the system is operated with a critical level (or level of alarm) $L_C$, which ensures a false positive rate no greater than  5\% (eq. \ref{EQ:1}), was obtained by applying the Currie equation, eq. \ref{EQ:2} \cite{knoll_CurieEq}:

\begin{equation}
  L_C=2.33\sigma_{Nb}\approx 53\,\textnormal{counts}
  \label{EQ:1}
\end{equation}
\begin{equation}
  N_D=4.65\sqrt{Nb}+2.71\approx 111\,\textnormal{counts}
   \label{EQ:2}
\end{equation}

The system total counts corresponding to the \emph{MDA} ($N_T$) is given by the sum of the background counts ($N_B$) and the corresponding net counts:

\begin{equation}
  N_T=N_D+N_B=111+540=651\,\textnormal{counts}
   \label{EQ:3}
\end{equation}

\begin{figure}[tb]
	\centering
		\includegraphics[width=0.5\columnwidth]{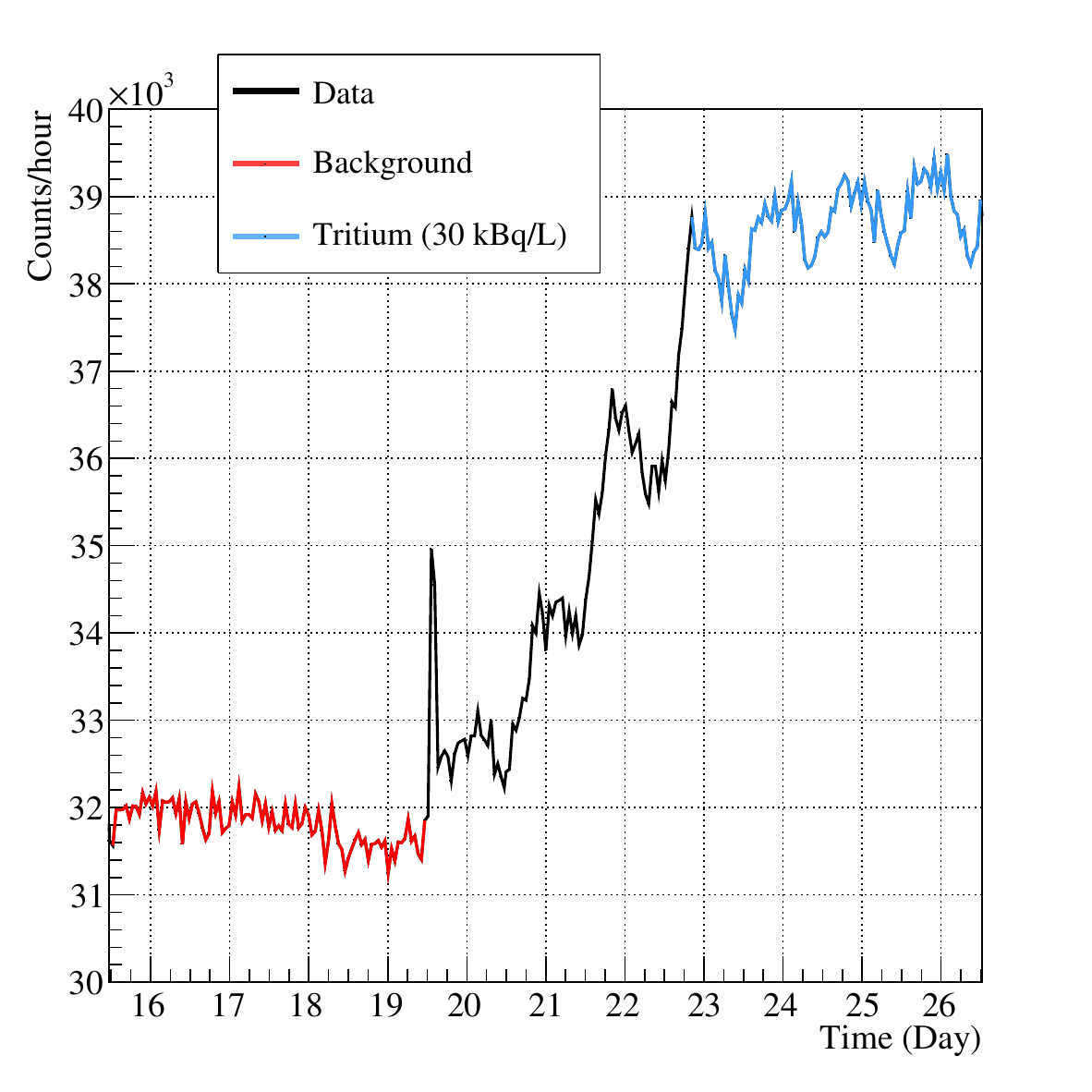}
	\caption{Counting rate (60\,min integration time) by increasing the tritium concentrations till 30\,kBq/L}
	\label{FIG:11}
\end{figure}

To experimentally measure the $MDA$, tritium liquid sources were slowly added to the water and recirculated in a closed water circuit using a small water pump. The tritium water activity was continuously increased until an averaged $N_T\approx$\,656\,counts/min was achieved. The blue lines and blue distribution in Figures \ref{FIG:10a} and \ref{FIG:10b}, respectively, present the acquired data with tritium. The final water activity was measured by a Quantulus system obtaining a value of 29.8$\pm$3.6\,kBq/L for the $MDA$.\\
As shown in \cite{Azevedo2020} the sensitivity can be increased by increasing the counting time. Thus, the data of Figure \ref{FIG:10a} were integrated over 60\,min being presented in Figure \ref{FIG:11}.  In this case a $\sigma_{Nb}=$225$\pm16$\,counts/min was obtained. By substituting this value in eq. \ref{EQ:2} a $N_D\approx$832\,counts/min is calculated. To obtain the $MDA$ for a 60\,min integration time ($MDA_{60min}$) we have assumed a linear relation between the background ($N_B=3,19\times 10^4$ for 0\,Bq/L) and the counting average of $3.87\times 10^4$ for a tritium activity of 30\,kBq/L which allows to extrapolate a $MDA_{60min}\approx$3.6$\pm$0.1\,kBq/L. A periodic oscillation coincident with the day cycle is observed after the  19\textsuperscript{th} day revealing a possible light leak due to the installation of the water closed circuit pump.

\begin{figure}[tb]
	\centering
		\includegraphics[width=0.5\columnwidth]{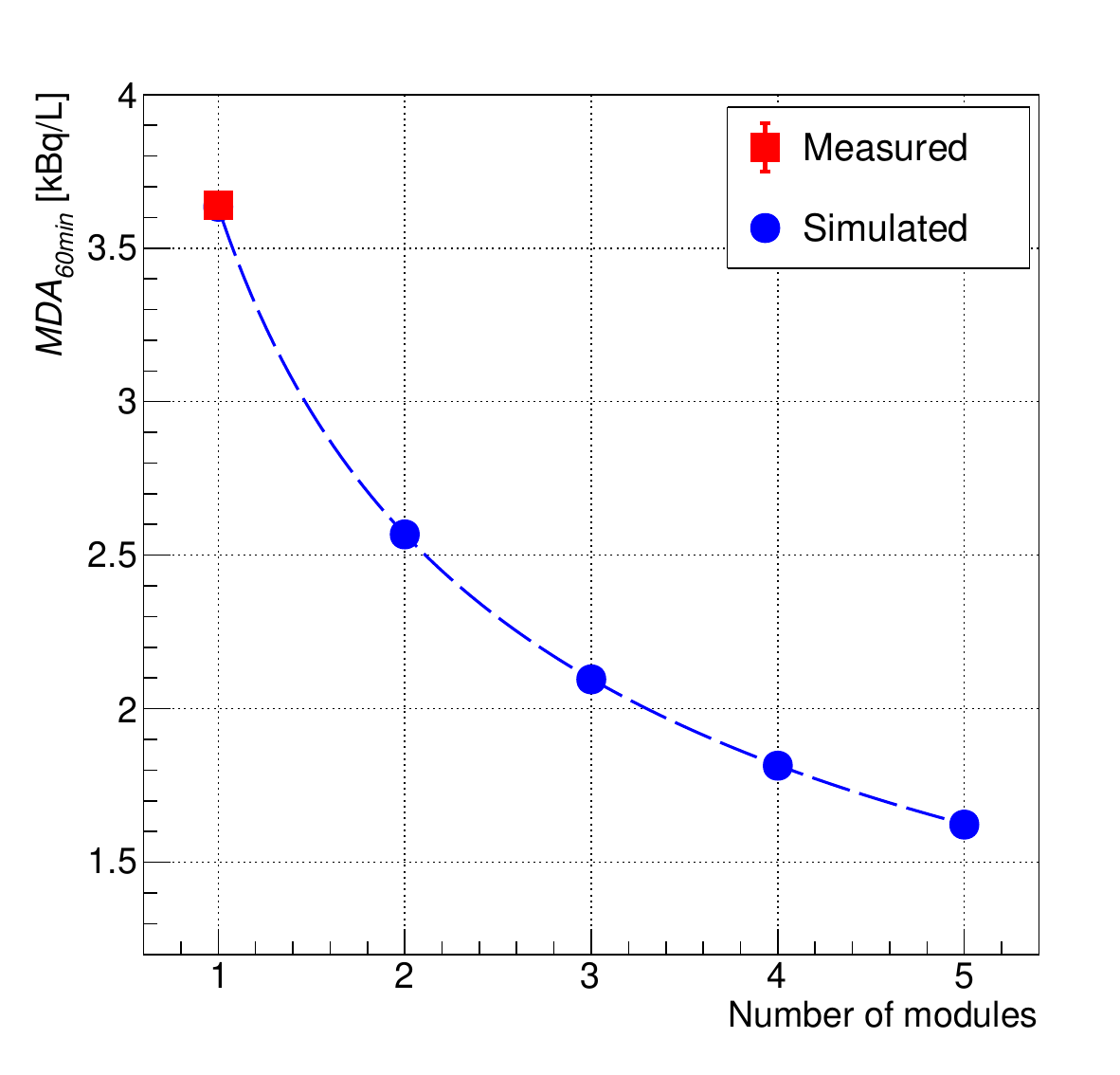}
	\caption{MDA (60\,min integration time) measured and the extrapolation for a detection system composed from 1 to 5 modules}
	\label{FIG:13}
\end{figure}

To evaluate the system performance by increasing the number of modules, an extrapolation was performed with the results being presented in Figure \ref{FIG:13}. The extrapolation was conceived by random sampling the distributions presented in Fig. \ref{FIG:10b}, both the background and the measured 30\,KBq/L, integrated over 60\,min and summed according to the number of modules. A total of 1.6 million events/hour were used. By including more modules in the extrapoled system the decrease on the $MDA_{60min}$ is observed. The dashed line is the result of the fit of eq. \ref{EQ:4} to the extrapolated data, where $K$ is a constant, $y$ and $x$ corresponds to the $MDA_{60min}$ and number of modules, respectively. From the extrapolation we concluded that adding more than 5 modules will still improve the system $MDA_{60min}$ but not significantly. To further improve the system sensitivity a background reduction is mandatory.

\begin{equation}
  y(x)=K\frac{1}{\sqrt{x}}
   \label{EQ:4}
\end{equation}

\subsection{Arrocampo installation}
\begin{figure}[hbt]
	\centering
		\includegraphics[width=0.5\columnwidth]{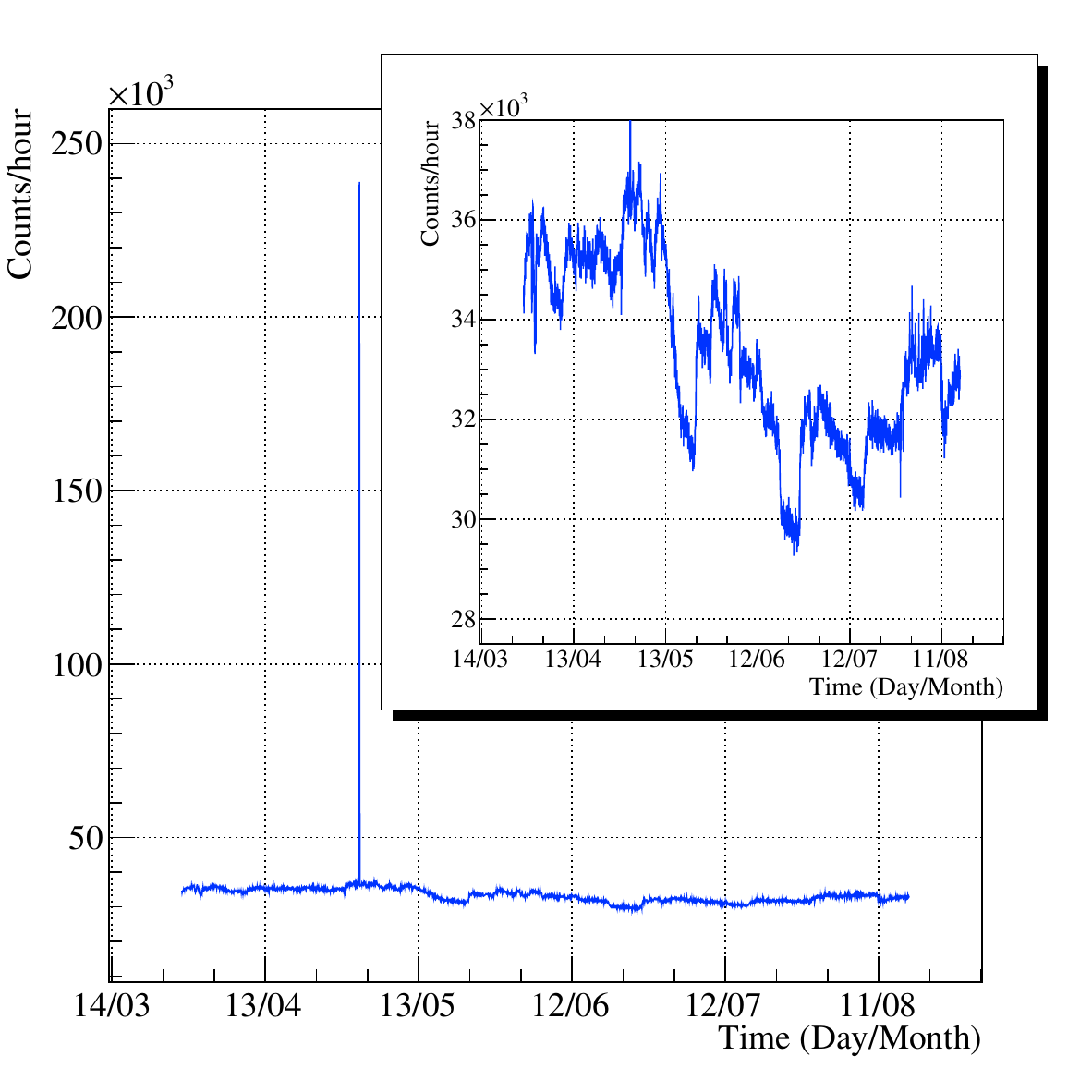}
	\caption{Counting rate of the prototype installed in Arrocampo facility using a 60\,min integration time. The figure presents the full scale, while the inset is a zoom-in on the data of interest.}
	\label{FIG:12}
\end{figure}

On the 27\textsuperscript{th} of March 2019 the prototype was moved from the laboratories of the University of Extremadura to the Arrocampo facility located in the close vicinity of the discharge channel from Arrocampo dam (Almaraz nuclear power plant) to the Tagus river. The facility owns a lead shield (5\,cm thickness) and a water de-ionizing system that produces water with a conductivity close to 10$\mu$S/cm \cite{corbacho}. The prototype was installed in order to perform a long term-measurement for the assessment of the detector stability. The acquired data, in a 60\,min integration time, are presented in Figure \ref{FIG:12}. The peak at the beginning of May was due to the removal of the lead shield roof to inspect the prototype and looking for possible water leaks. An increase of a factor 6.7 in the counting rate was measured by just removing the lead roof. For better visualization, the data were zoomed-in and presented in the inset of the figure. During the full measurement period, it is possible to observe an amplitude variation corresponding to 6 times the $MDA_{60min}$, whose value was calculated in the previous subsection. A careful analysis of the full logged data allowed us to identify the probable sources for this variation: random electronic noise with possible origin on the powerful water pumps of the purification system and fluctuations in the DAC voltages (used as input for the comparators) were also registered. The instabilities in the DAC voltage levels and the random noise are time compatible with the sudden variations observed in the coincidence counter.

\section{Future improvements}
During the present work, several issues were identified and the corresponding solutions proposed. First, the PMTs false coincidence rate must be decreased by changing the PMTs bias from negative to positive high voltage. With this technique used in photon counting applications, a decrease of the thermal emission from the photocathode is foreseen and consequently a decrease of the false coincidence rate \cite{knoll_PMTS_2}. Moreover, the choice of low radioactivity glass for the PMTs will be mandatory for the achievement of the 100\,Bq/L sensitivity. The installation of a cosmic veto discriminator is needed for the background reduction, which is the main background source in the present conditions. A better grounding to avoid random noise and new electronics boards will be produced with industry grade. With the inclusion of more modules to increase the sensitivity, the counting cannot be performed by a Raspberry Pi. A counter board based on FPGA will be required to ensure a fast and reliable counting. The new modules must be designed to increase the light tightness, for example in the connectors and electric feed-throughs located at the PVC endcap. 

\section{Conclusions}

In this work we have presented the developments of a prototype module for the construction of a real-time tritium-in-water monitor with a sensitivity of 100\,Bq/L. With a single module we have measured a $MDA$ of 30\,kBq/L with an integration time of 1\,min. If considering a 60\,min time integration we extrapolated a remarkable $MDA$ value of 3.6\,kBq/L. The obtained values are already in the range of environmental surveillance and nuclear power plant pre-alert system considering a single only module. By using a system composed by 5 modules and a 60 min counting time, a $MDA$ close to 1.5\,kBq/L limited by the background level. 
During this work, several issues were found and the solutions addressed. The main source of system instabilities was related to light-tightness and random electronic noise.

\acknowledgments

This work was supported by the INTERREG-SUDOE program through the project TRITIUM - SOE1/P4/E0214.
C.D.R. Azevedo was supported by Portuguese national funds (OE), through FCT - Funda\c{c}\~{a}o para a Ci\^{e}ncia e a Tecnologia , I.P., in the scope of the Law 57/2017, of July 19.
Part of this work was also developed within the scope of the project I3N, UIDB/50025/2020 \& UIDP/50025/2020, financed by national funds through the FCT/MEC.


\end{document}